\documentclass[%
 reprint,
 amsmath,amssymb,
 aps,
]{revtex4-2}

\usepackage{graphicx}
\usepackage{dcolumn}
\usepackage{bm}
\usepackage{hyperref}
\hypersetup{
  colorlinks   = true, 
  urlcolor     = blue, 
  linkcolor    = black, 
  citecolor   = blue 
}
\usepackage{amsmath}  
\DeclareMathOperator{\Var}{Var}
\usepackage{amsfonts} 
\usepackage{graphicx} 
\usepackage{physics}
\usepackage{qcircuit}

\begin{document}

\preprint{APS/123-QED}

\title{Modeling Quantum Enhanced Sensing on a Quantum Computer}

\author{Cindy Tran}
\email{cindyt@mit.edu}
\affiliation{Massachusetts Institute of Technology, Cambridge, MA 02139}
\author{Tanaporn Na Narong}
\author{Eric S. Cooper}
\email{escooper@stanford.edu}
\affiliation{Department of Physics, Stanford University, Stanford, CA 94305}

\date{\today}

\begin{abstract}
Quantum computers allow for direct simulation of the quantum interference and entanglement used in modern interferometry experiments with applications ranging from biological sensing to gravitational wave detection. Inspired by recent developments in quantum sensing at the Laser Interferometer Gravitational-wave Observatory (LIGO), here we present two quantum circuit models that demonstrate the role of quantum mechanics and entanglement in modern precision sensors. We implemented these quantum circuits on IBM quantum processors, using a single qubit to represent independent photons traveling through the LIGO interferometer and two entangled qubits to illustrate the improved sensitivity that LIGO has achieved by using non-classical states of light. The one-qubit interferometer illustrates how projection noise in the measurement of independent photons corresponds to phase sensitivity at the standard quantum limit. In the presence of technical noise on a real quantum computer, this interferometer achieves the sensitivity of 11\% above the standard quantum limit. The two-qubit interferometer demonstrates how entanglement circumvents the limits imposed by the quantum shot noise, achieving the phase sensitivity 17\% below the standard quantum limit. These experiments illustrate the role that quantum mechanics plays in setting new records for precision measurements on platforms like LIGO. The experiments are broadly accessible, remotely executable activities that are well suited for introducing undergraduate students to quantum computation, error propagation, and quantum sensing on real quantum hardware.
\end{abstract}

\maketitle


\section{Introduction} 
Quantum systems at the forefront of modern research increasingly utilize entanglement for goals ranging from precision sensing to quantum simulation and computation. Prominently, quantum computers employ entanglement to push beyond the capabilities of classical computers, providing a new set of tools that promise to solve complex problems across multiple fields. On the way to more advanced research goals, first generation quantum computers also serve as educational tools. The IBM Quantum Experience, for example, is an online quantum computing platform that provides the opportunity for students to interact with real quantum hardware~\cite{brody2021spin, johnstuna2021understanding, kain2021searching, encinar2021digital}. Here we use the IBM Quantum Experience to teach students about quantum metrology, one of the first areas of quantum science where entanglement has found practical applications.

Applications for entanglement in quantum metrology include spin squeezed clocks\cite{clocks}, enhanced biological imaging\cite{bio_imaging}, remote sensing with EPR states~\cite{remote_sensing}, and the detection of gravitational waves~\cite{abbot2016observation, tse2019quantum, georgescu2020o3}. Our focus for this paper will be on the particularly charismatic example of LIGO~\cite{cooper2021interactive}, which made the first detection of gravitational waves in 2015, realizing a century-old prediction by Einstein~\cite{abbot2016observation}. At the heart of LIGO is the most sensitive interferometer ever built, whose sensitivity has recently been enhanced by quantum squeezing of light~\cite{tse2019quantum}. LIGO clearly illustrates the quantum nature of interferometers and the role that entanglement can play in modern metrology. 

Here, we describe how to use the IBM Quantum Experience to understand the quantum physics underpinning discoveries at LIGO. In Sec.~\ref{Sec:LIGO-to-circuit}, we construct a single-qubit quantum circuit representation of the LIGO interferometer by analyzing the behavior of a photon traveling through the interferometer. In Sec.~\ref{Sec:1-qubit-behavior}, we present two representations of this quantum circuit - matrices and Bloch spheres - which provide precise mathematical intuition for the behavior of the quantum circuit. We test predictions made by these representations  against the results of experiments performed on the IBM Quantum Experience in Sec.~\ref{Sec:1-qubit-result}. These results allow us to model the nature of quantum noise in the interferometer. We present a calibrated measurement of the sensitivity of the interferometer in Sec.~\ref{Sec:1-qubit-noise}. 

The tools developed through analysis of the single-qubit interferometer circuit allow us to efficiently assess the enhancement that can be provided by quantum entanglement. In Sec.~\ref{Sec:2-qubit-result}, we present an interferometer circuit with entanglement between two qubits: the simplest possible example of quantum-enhanced metrology. We model the expected behavior of this circuit using matrix mechanics and visualize the expected dynamics using spin Wigner functions~\cite{dowling1994wigner} before implementing the circuit on a quantum computer. We show in Sec.~\ref{Sec:2-qubit-noise} that entanglement reduces the noise of this interferometer below the standard quantum limit, even in the presence of experimental imperfections. We present our conclusions in Sec.~\ref{Sec:conclusions}.

\section{LIGO as a quantum circuit}\label{Sec:LIGO-to-circuit}

\begin{figure*}[ht]
    \centering
    \includegraphics[width=0.75\textwidth]{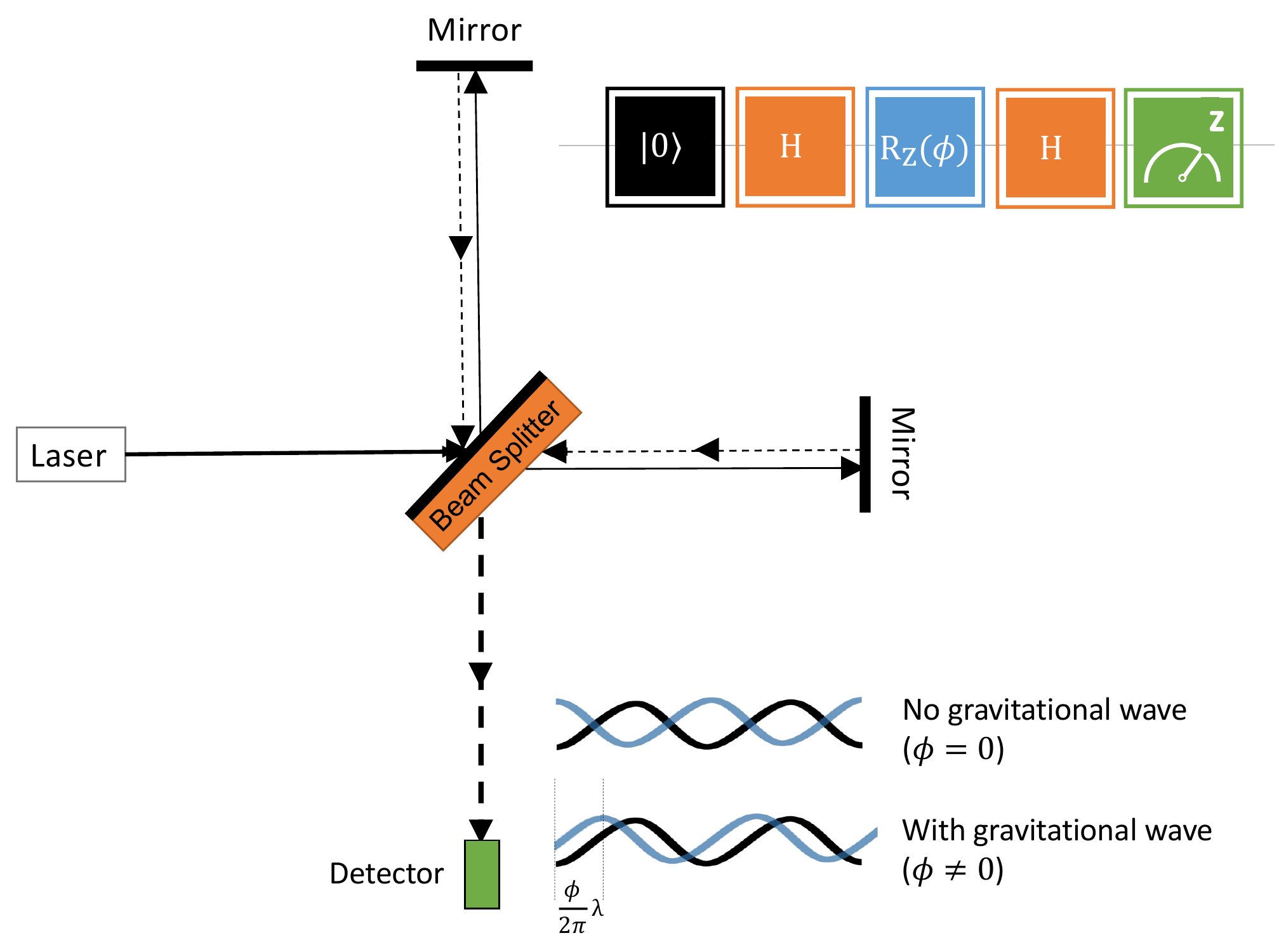}
    \caption{A simplified version of the LIGO interferometer is presented alongside an equivalent quantum circuit for a single photon. Photons traveling parallel to the laser are represented by the qubit state $|0\rangle$ while photons traveling perpendicularly to the laser are represented by the qubit state $|1\rangle$. A photon emitted from the laser reaches the beam splitter, which produces a superposition state of the transmission ($|0\rangle$) and reflection ($|1\rangle$), analogous to a Hadamard gate (H). While in this superposition, the photon travels along both of the 4-km long perpendicular arms, accumulating a phase shift $\phi$ that depends on the relative length of the two arms. This corresponds to an $R_Z(\phi)$ rotation gate, which applies the phase shift $\phi$ to a qubit superposition. After recombination on the beam splitter, the phase shift determines the interference pattern on the detector. For destructive interference ($\phi = 0$), the beam splitter returns the photon to the laser and quantum circuit measurement yields a result of $|0\rangle$. For other phase shifts ($\phi \neq 0$), the beam splitter is more likely to direct the photon onto the detector, corresponding to a circuit measurement of $|1\rangle$. The number of photons measured at the detector thus depends on the relative length of the arms and is sensitive to passing gravitational waves.}
    \label{fig: Interferometer optics in relation to quantum circuit}
    \label{fig:LIGO-circuit}
\end{figure*}

LIGO utilizes laser interferometry to measure the microscopic spacetime distortions caused by gravitational waves. As shown in Fig.~\ref{fig:LIGO-circuit}, these distortions change the relative length of LIGO's two arms and thus change the interference signal on the detector. The resulting signals are so small that measurements in the frequency bands relevant to the final moments of black hole and neutron star mergers are limited by the inherent quantum shot noise of the laser used in the interferometer~\cite{abbott2020noise}. Thus there is value in incorporating quantum mechanics into the description of the interferometer, going beyond the classical framework~\cite{cooper2021interactive} of understanding interferometers with electromagnetic waves. With only one qubit, a quantum circuit model can capture all the salient features of a classical model and demonstrate the effects of quantum projection noise. Repeatedly executing this circuit simulates measurements of the independent photons that make up a coherent laser beam. The quantum circuit model also provides a scalable framework that can incorporate entanglement, which is necessary for quantum enhanced measurement~\cite{caves_2010_quantumcircuit}.



To construct a quantum circuit model of LIGO, we pare the complicated optical system down to a Michelson interferometer, consisting of only a laser, a beam splitter and two mirrors. This geometry, depicted in Fig.~\ref{fig:LIGO-circuit}, captures the essential physics of the system and is often used to illustrate the operation of LIGO~\cite{cooper2021interactive,operation}. We map the two directions that a photon travels in the interferometer to the two states of a single qubit in the quantum circuit. A photon traveling parallel to the laser corresponds to the qubit state $|0\rangle$, and a photon traveling in the perpendicular direction corresponds to the state $|1\rangle$. We deduce the gates that make up the quantum circuit by analyzing the behavior of each of the optical elements that a photon encounters while traveling through the interferometer. This procedure is similar to that used by Nielsen and Chuang to construct a Mach–Zehnder interferometer in the context of dual-rail quantum computation (Section 7.4.2)~\cite{nielsen2010quantum}.

A photon in the interferometer is initially emitted by a monochromatic laser source and travels towards the beam splitter. The initial direction of the photon is represented by the quantum state $|0\rangle$. At the beam splitter, the photon has 50\% chance of reflection. This creates an equal superposition of the photon continuing to travel in the same direction (quantum state $|0\rangle$) and being reflected into the perpendicular direction (quantum state $|1\rangle$), which precisely maps onto the action of a Hadamard gate on a single qubit.  After the beam splitter, each state accumulates a phase that is proportional to the distance traveled in the respective arm. A difference in length of the two arms, which can be caused by a gravitational wave, introduces a phase difference $\phi$ between the two states. This is analogous to the action of an $R_Z(\phi)$ gate. After retro-reflection at the end of each arm, the photon returns for a second pass through the beam splitter, which is again represented by a Hadamard gate. After the beam splitter, photons in the state corresponding to $|1\rangle$ are measured on the detector, constituting a measurement in the qubit's computational ($z$) basis.

Putting together the steps of the photon's journey, the full quantum circuit that represents the LIGO interferometer is 
\begin{equation}\label{1-qubit-circuit}
    \Qcircuit @C=1.0em @R=0.2em @!R {
	 	\lstick{ {q}_{0} :  } & \gate{\left|0\right\rangle} & \gate{\mathrm{H}} & \gate{\mathrm{R}_\mathrm{Z}\,\mathrm{(}\mathrm{\phi}\mathrm{)}} & \gate{\mathrm{H}} & \meter & \qw & \qw\\
	 	\lstick{c:} & \lstick{/_{_{1}}} \cw & \cw & \cw & \cw & \dstick{_{_{0}}} \cw \cwx[-1] & \cw & \cw\\
	 }
\end{equation}
This circuit allows us to directly simulate the behavior of a Michelson interferometer on a quantum computer. 

\section{Expected Behavior of the Single-Qubit Circuit}\label{Sec:1-qubit-behavior}
In order to formulate a precise quantum picture of a Michelson interferometer like LIGO, we require a mathematical framework that can predict the behavior of the quantum circuit presented in the prior section. Quantum matrix mechanics provides this framework, but is mathematically removed from a physical or graphical interpretation. We thus present the Bloch sphere representation alongside the matrix representation to help visualize the action of the quantum gates on the qubit. For a single qubit, both complementary representations independently provide a precise accounting of the quantum dynamics.

The matrix notation for quantum states represents a qubit as a $2\times1$ vector whose entries are complex numbers. Quantum gates are represented by $2\times2$ matrices, which act on the state vector through matrix multiplication. Following the convention of Nielsen and Chuang~\cite{nielsen2010quantum}, the two basis states $|0\rangle$ and $|1\rangle$ are given by
\begin{equation}
    |0\rangle = \begin{bmatrix}  
        1 \\
        0
        \end{bmatrix} \text{and }
    |1\rangle = \begin{bmatrix}  
    0 \\
    1
    \end{bmatrix}.
\end{equation}
An arbitrary superposition of $|0\rangle$ and $|1\rangle$ is given by $|\psi\rangle =  c_0|0\rangle + c_1|1\rangle$ and can be written as the vector
\begin{equation}
    |\psi\rangle = \begin{bmatrix}  
        c_0 \\
        c_1
        \end{bmatrix}.
\end{equation}
The squared magnitudes of the coefficients $|c_0|^2$ and $|c_1|^2$ give the probabilities of measuring the qubit in the $|0\rangle$ and $|1\rangle$ states, respectively. In the context of the interferometer, only photons in the $|1\rangle$ state (traveling perpendicularly to the laser output beam) reach the detector whereas photons in the $|0\rangle$ state (traveling along the laser output beam) return to the laser without being detected. The probabilities $p_1 = |c_1|^2$ and $p_0 = |c_0|^2$ hence respectively correspond to the probabilities of detecting and not detecting each photon. Information about the quantum superposition is encapsulated in the relative phase $\phi = \phi_1 - \phi_0$ between the two complex coefficients, where $\phi_0$ and $\phi_1$ are defined by the expression of the complex coefficients in polar form: $c_0 = |c_0|e^{i\phi_0}$ and $c_1 = |c_1|e^{i\phi_1}$.

\begin{figure*}[ht]
    \centering
    \includegraphics[width=0.75\textwidth]{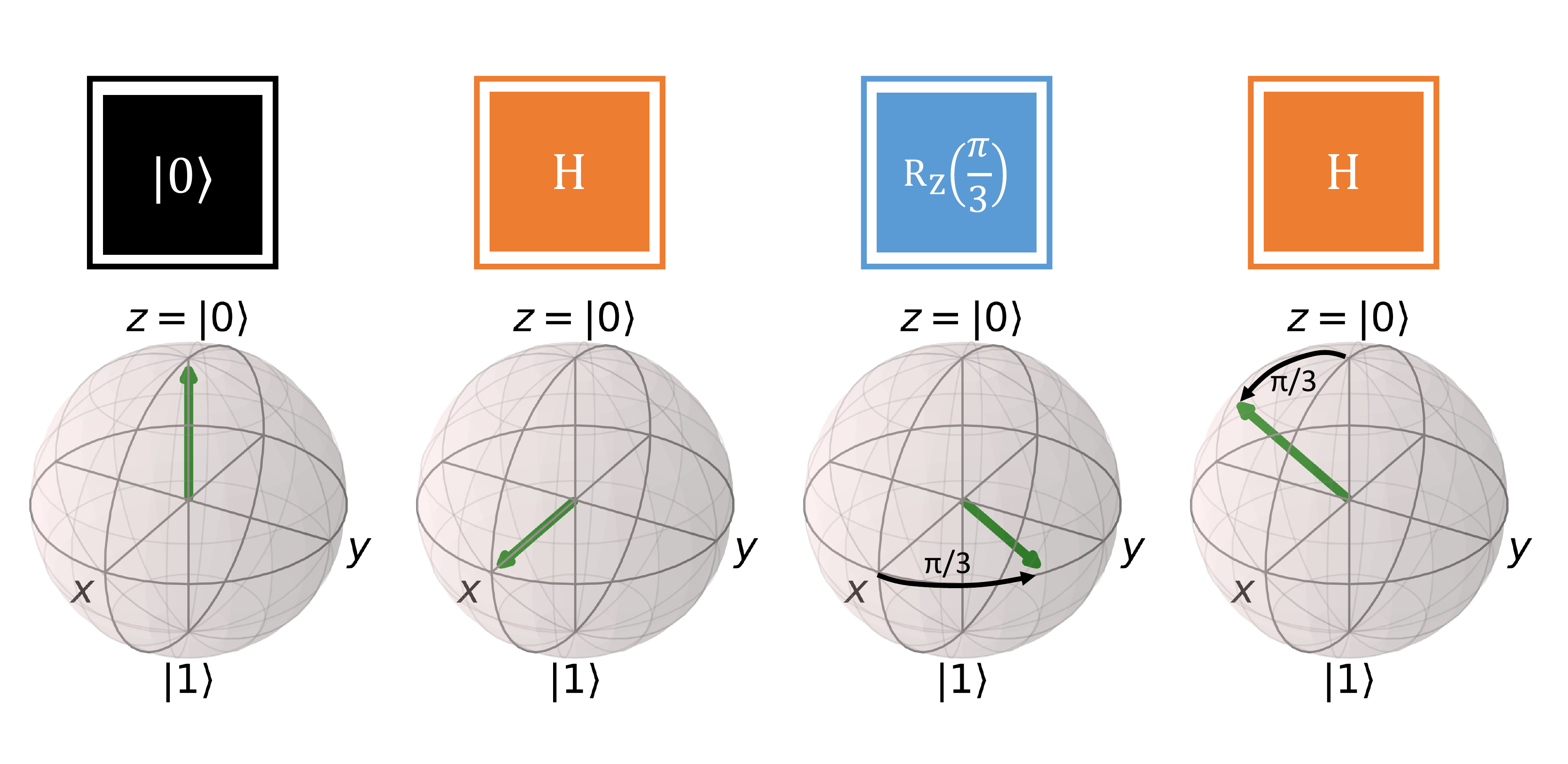}
    \caption{Visualization on the Bloch sphere of the quantum state of the single-qubit circuit after each gate is applied. The arrow shown on each sphere represents the quantum state $|\psi\rangle = c_0|0\rangle + c_1|1\rangle.$ The $z$ coordinate is the relative probability $p_0 - p_1$ of measuring $|0\rangle$ or $|1\rangle$ while the azimuthal angle is the relative phase $\phi$ between the complex coefficients $c_0$ and $c_1$. The initial state $|0\rangle$ is a vector pointing up. The Hadamard gate rotates the state by $\pi$ about the $z$-axis, then by $\pi/2$ about the $y$-axis. The $R_Z(\phi)$ gate rotates the state about the $z$-axis, with $\phi=\pi/3$ chosen for illustration. The final Hadamard gate maps $x$-axis back onto the z-axis, which allows $\phi$ to be measured.}
    \label{fig:Bloch-sphere-1-qubit}
\end{figure*}

The quantum state $|\psi\rangle$ of a qubit can also be represented on the Bloch sphere as seen in Fig.~\ref{fig:Bloch-sphere-1-qubit}. The Bloch sphere is a unit sphere where a vector from the origin to the surface of the sphere represents a single-qubit state. This vector is known as the Bloch vector. The state $|0\rangle$ is represented by a spin pointing up ($+z$) and $|1\rangle$ is represented by a spin pointing down ($-z$). For superpositions of $|0\rangle$ and $|1\rangle$, the $z$ coordinate is given by the relative probability of measuring each of the two states, $p_0 - p_1$. The angle in the $xy$-plane is given by the quantum phase $\phi$, with $\phi=0$ corresponding to the $x$-axis. Quantum gates generate rotations of the state vector on the Bloch sphere.

During the sequence, the qubit is initialized in the state $|0\rangle$, represented by a vector pointing up on the Bloch sphere in Fig.~\ref{fig:Bloch-sphere-1-qubit}a. The first gate applied to this initial state is a Hadamard gate $H$. The matrix form of $H$ is
\begin{equation}
    H = \frac{1}{\sqrt{2}}\begin{bmatrix}  
        1 & 1 \\
        1 & -1
        \end{bmatrix}.
\end{equation}
On the Bloch sphere, the Hadamard gate first rotates the state by $\pi$ about $z$ and then by $\pi/2$ about $y$. This transforms the initial state $|0\rangle$ into an equal superposition of $|0\rangle$ and $|1\rangle$ with a relative phase of 0, 
\begin{equation}
    |\psi\rangle = \frac{1}{\sqrt{2}}\begin{bmatrix}  
        1  \\
        1 
        \end{bmatrix}.
\end{equation}
This state vector lies along the $x$-axis of the Bloch sphere. 

 Following the Hadamard gate, the $R_Z(\phi)$ gate rotates the state about the $z$-axis by an angle $\phi$. As an example, a rotation by $\phi = \pi/3$ is depicted in Fig.~\ref{fig:Bloch-sphere-1-qubit}. In the matrix formulation, the $R_Z(\phi)$ gate introduces a relative phase of $\phi$ between the two complex coefficients $c_0$ and $c_1$ via the action of the matrix
        \begin{equation}
    R_Z(\phi) = \begin{bmatrix}  
        e^{-i\phi/2} & 0 \\
        0 & e^{i\phi/2}
        \end{bmatrix}.
\end{equation}
For an interferometer like LIGO, this relative phase $\phi$ is induced by a path length difference of $\Delta L = \phi\lambda/2\pi$ where $\lambda$ is the wavelength of light used in the interferometer. Because the quantum state changes based on the value of $\phi$, the detector is sensitive to the effects of gravitational waves.


In order to map the phase accumulated due to the $R_Z(\phi)$ back into the measurement basis ($z$), we apply a final Hadamard gate. This gate maps the $x$-axis back to the $z$-axis so that phase accumulations in the $xy$-plane manifest as changes in expectation value for measurements in the standard basis of $|0\rangle$ and $|1\rangle$. Mathematically, we compute the resulting state of the quantum circuit by multiplying together all of the gates so that the final state is $|\psi\rangle = HR_Z(\phi)H|0\rangle$, which evaluates to 
\begin{equation}\label{eq:1-qubit_finalstate}
|\psi\rangle = 
\frac{1}{2}\begin{bmatrix}
e^{-i\phi/2} + e^{i\phi/2}\\
e^{-i\phi/2} - e^{i\phi/2}
\end{bmatrix}
=\begin{bmatrix}
\cos{(\phi/2)}\\
-i\sin{(\phi/2)}
\end{bmatrix}.
\end{equation}
Here, the probabilities of measuring the final state of the circuit as $|0\rangle$ and $|1\rangle$ are given by $p_0 = \cos^2(\phi/2)$ and $p_1 = \sin^2(\phi/2)$, respectively.

We summarize the result of this measurement by defining the polarization operator $\hat{P}$ such that the polarization is equal to -1 if we measure $|0\rangle$ and equal to 1 if we measure $|1\rangle$. Up to the overall sign, this is the $z$ coordinate of the Bloch vector. In matrix form, the polarization operator is
\begin{equation}
    \hat{P} = \begin{bmatrix}  
        -1 & 0 \\
        0 & 1
    \end{bmatrix}.
\end{equation}
For the final state $|\psi\rangle$ from Eq.~\eqref{eq:1-qubit_finalstate}, the expectation value of the polarization operator is
\begin{equation}\label{eq:polarization-definition}
    \langle \psi | \hat{P} |\psi\rangle = p_1 - p_0 = \frac{1}{2}\sin^2(\phi/2) - \frac{1}{2}\cos^2(\phi/2) = -\cos{\phi}
\end{equation}
This sinusoidal dependence on $\phi$ is consistent with the behavior of the $z$-coordinate of the final state visualized in Fig.~\ref{fig:Bloch-sphere-1-qubit}, where the final state's Bloch vector makes an angle $\phi$ with respect to the $z$ axis. In the context of LIGO, this dependence on $\phi$ facilitates sensitivity to signals from gravitational waves. The mapping between the LIGO interferometer and the quantum circuit allows us to simulate the detection of gravitational waves on the quantum computer and better understand the impacts of both fundamental and technical noise sources.

\section{Implementing on a quantum computer}\label{Sec:1-qubit-result}
We implemented the quantum circuit model for LIGO on the IBM Quantum Experience. Quantum circuits were initially constructed manually via the circuit composer interface, which allows quick prototyping without the need for any programming experience. To scale up the number experimental trials and systematically scan through experimental parameters, we used the Qiskit interface, which allows users to program quantum circuits in Python~\cite{asfaw2020}. The code for interfacing with the quantum computers was adapted from existing examples utilizing IBM Quantum Experience~\cite{brody2021spin, kain2021searching, asfaw2020}. The code used for all experiments in this paper is available online~\cite{github}.

Experiments were performed on both the quantum computer ibmq\_manila and the state vector simulator ibmq\_qasm\_simulator. The quantum computer ibmq\_manila was chosen because, with a quantum volume of 32, it had the largest quantum volume among freely accessible quantum computers~\cite{IBM_compute}. The quantum volume is IBM's preferred metric for its quantum computers and is a proxy for overall error rates; it measures the size of the largest quantum circuit that can be executed with acceptable fidelity~\cite{quantum_volume}. The simulator ibmq\_qasm\_simulator explicitly computes the final qubit state and then samples outcomes from the corresponding probability distribution. This provides a reference point for the effects of finite quantum volume and allows us to distinguish the effects of statistical noise from technical noise sources on the quantum computer.

\begin{figure}[tp]
\centering
\includegraphics[width=0.47\textwidth]{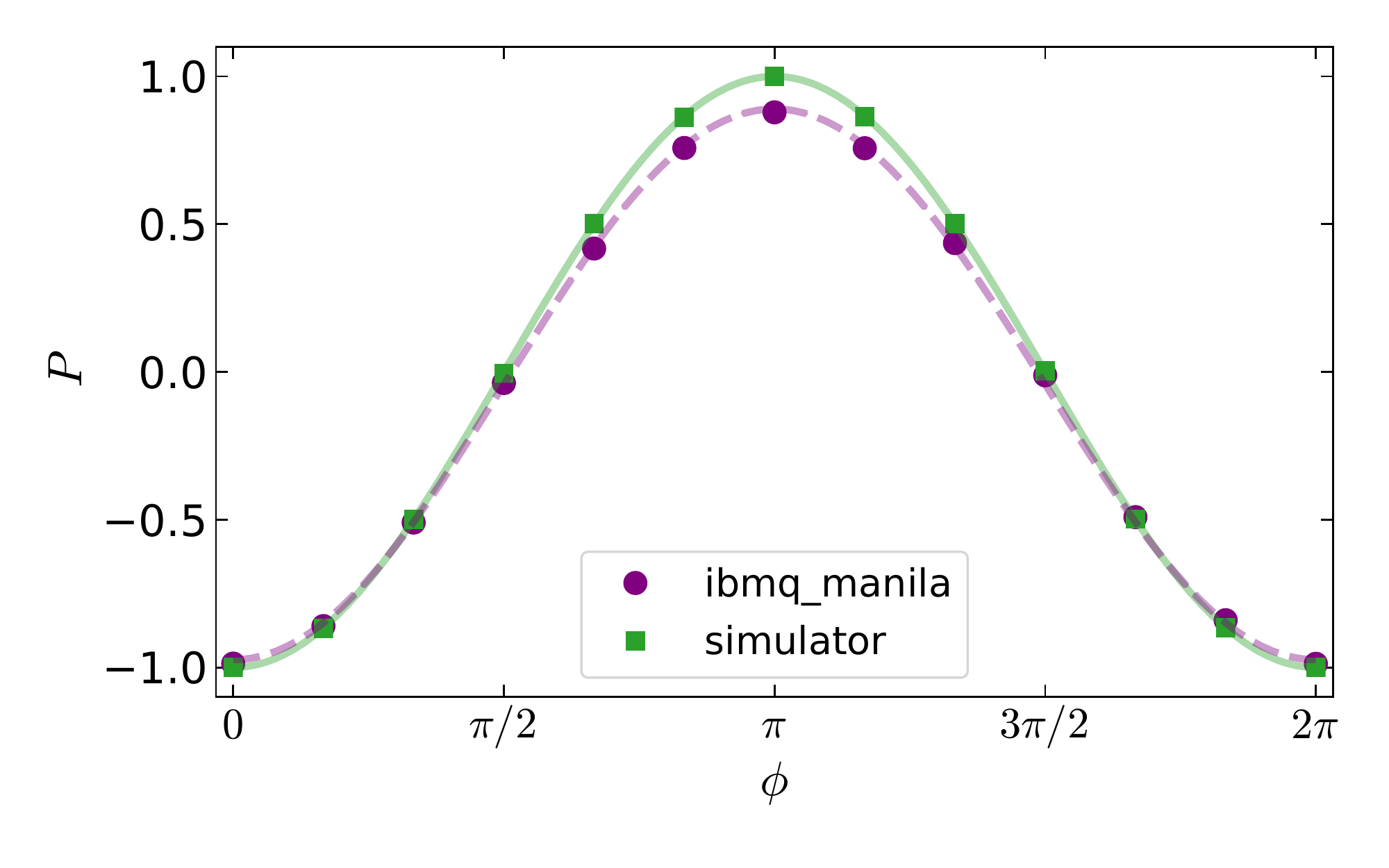}
\caption{Measured polarization $P = p_1 - p_0$ as a function of $\phi$, computed from the data taken on the quantum computer (purple circle) and the simulator (green square) at 13 angles between 0 and $2\pi$, along with the best-fitted curves. Error bars are smaller than the data points. The curves are fitted to the data with amplitude and offset as free parameters.}
\label{fig:1-qubit_polarization}
\end{figure}

Figure~\ref{fig:1-qubit_polarization} shows plots of polarization $P$ as a function of $\phi$, measured by running the single-qubit circuit from Eq.~\eqref{1-qubit-circuit} on both the quantum computer ibmq\_manila and the simulator ibmq\_qasm\_simulator. Each data point is obtained by repeatedly executing the quantum circuit and tallying the measurement outcomes 0 and 1 from each shot. A single trial consists of 1024 shots from which we calculate the proportions of shots $p_0$ and $p_1$ with each measurement outcome. Each data point shown is the average polarization $P = p_1 - p_0$ from 5 trials conducted at a fixed value of $\phi$. The variability among the 5 trials is smaller than the size of the data points.

To analyze the relationship between $P$ and $\phi$, we fit a cosine curve to the data, with amplitude and vertical offset as free parameters. The fit to the simulator data has an amplitude of 0.999(1) and an offset of -0.0001(7), which are consistent with the expected values of 1 and 0 to within the standard errors reported by the fitting routine (shown in parentheses). This affirms the validity of the fitting routine. 

The fit to the experimental data from the quantum computer has an amplitude of 0.932(5) and an offset of -0.042(4). The amplitude is reduced from the expected value, mainly due to deviations from theory at positive values of polarization. The majority of these deviations can be explained by readout errors. The readout errors on ibmq\_manila  vary between qubits and over time, but the overall error rate is typically reported between 2 and 3 percent~\cite{IBM_compute}, which corresponds to 4-6\% reductions in the amplitude of the polarization. The offset can be explained by unequal rates of the two types of readout error, where the dominant source of error is relaxation from the excited state $|1\rangle$ to the ground state $|0\rangle$ during measurement~\cite{IBM_compute}. IBM reports an error rates of $2\times10^{-4}$ for the main component of Hadamard gate, the $\sqrt{X}$ gate. This error rate is approximately one hundred times smaller than the readout error rates and thus a negligible contribution to the overall error. Additional errors can come from the $R_z$ gate, whose error rate is not reported by IBM.

These results verify that measurements of the single-qubit interferometer circuit depend on the interferometer phase $\phi$ as expected after incorporating corrections due to dissipation in the real quantum system. Dissipation primarily manifests as readout errors that reduce the amplitude of the polarization curve. The corrections to the curve are important to note as we next seek to invert the relationship and extract the phase of the interferometer from measurement outcomes.

\section{Shot noise and the Measurement of Gravitational Waves}\label{Sec:1-qubit-noise}
The sensitivity of an interferometer is defined by the smallest phase shift that can be reliably detected in the measured output of the interferometer. In general, this phase shift is a change from an initial interferometer phase $\phi_0$ to a new interferometer phase $\phi = \phi_0 + \delta\phi$. At LIGO, $\delta\phi$ is induced by gravitational waves and $\phi_0$ corresponds to the initial phase in the absence of gravitational waves. On the quantum computer we can choose $\delta\phi$ and $\phi_0$ arbitrarily to illustrate how shot noise limits the precision with which an interferometer can measure phase shifts.

\begin{figure}[tp]
\centering
\includegraphics[width=0.47\textwidth]{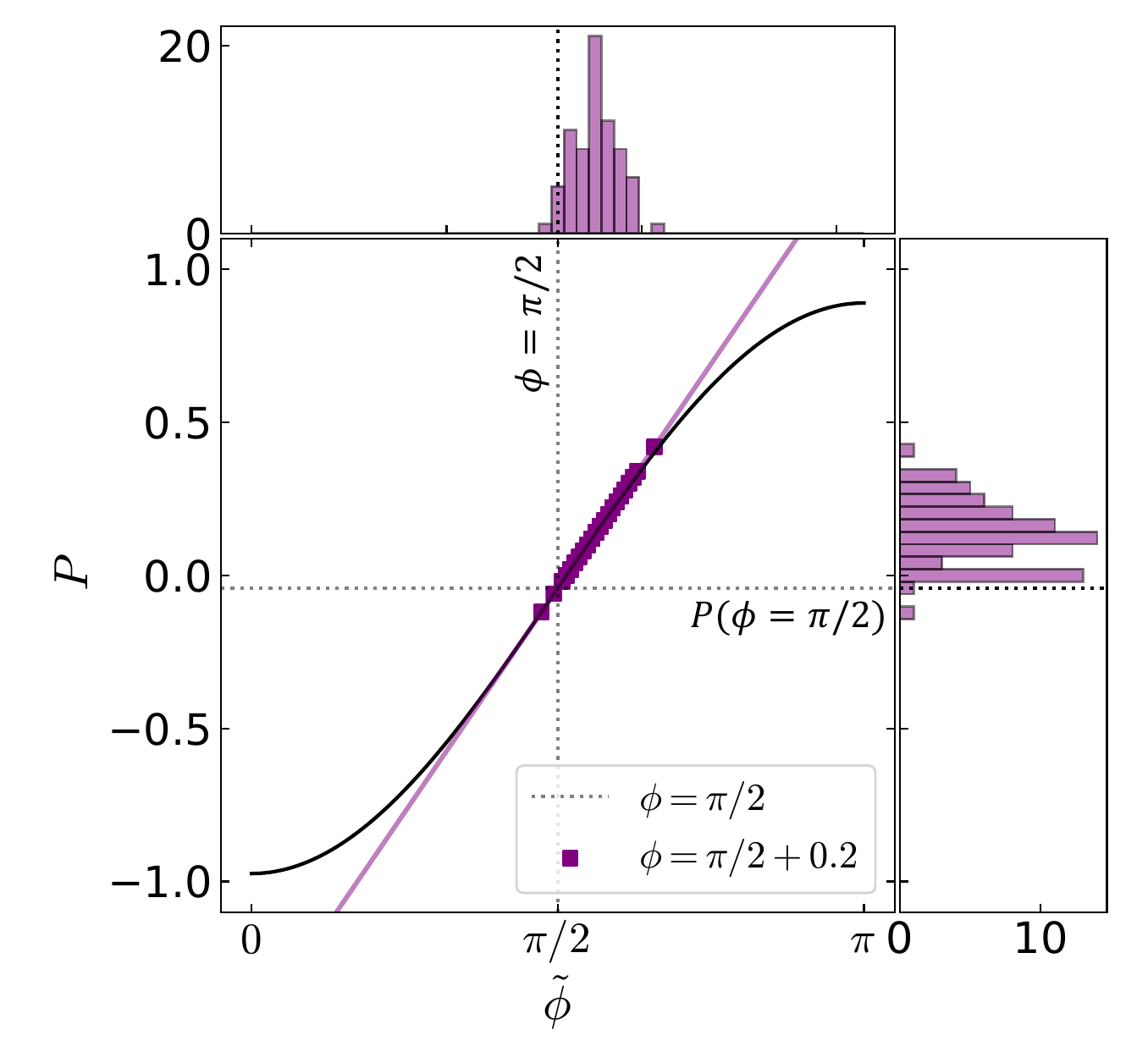}
\caption{Polarization measurements at $\phi = \pi/2 + 0.2$ (right histogram) and the corresponding inferred phases $\tilde{\phi}$ (top histogram) obtained by inverting the linear approximation (purple solid line) of the best-fitted $P$ vs $\phi$ cosine curve (black solid curve) at $\phi_0 = \pi/2$. The vertical and horizontal dotted lines represent the original phase value $\phi_0 = \pi/2$ and the corresponding polarization $P(\phi = \pi/2)$, respectively. The black curve is the ibmq\_manila cosine fit shown earlier in Fig.~\ref{fig:1-qubit_polarization}. The data were taken from 75 sequential polarization measurements, each consisting of $N=100$ shots.}
\label{fig:1-qubit-sensitivity}
\end{figure}

Since $\delta\phi$ is small, we can usefully approximate the cosine relationship between polarization and $\phi$ as locally linear, as depicted in the center panel of Fig.~\ref{fig:1-qubit-sensitivity}. Using this linear relationship, polarization measurements can be used to produce an estimate $\tilde{\phi}$ of the phase of the interferometer $\phi = \phi_0 + \delta\phi$. Error propagates from measurements of the polarization to the inferred values of $\tilde{\phi}$ with a scale factor given by the slope $m$ of the line, where $m=\sin(\phi_0)$ for an ideal interferometer. We can express the standard error of the inferred phase $\tilde{\phi}$ in terms of the standard deviation of polarization measurement as
\begin{equation}
    \sigma_{\tilde{\phi}} = \frac{1}{m}\sigma_{P}.
    \label{eq:noise_slope}
\end{equation}

Using this expression for the standard deviation of the inferred phase, we can determine theoretically how the precision of the interferometer should scale with number of qubits $N$ that are included in the measurement of average polarization $P$ from a single trial. Here $P = \frac{1}{N}\sum_{i=1}^N P_i$, where $P_i$ represents the measurement of an individual qubit. We can compute $\sigma_{P} = \sqrt{\Var(P)}$ by determining the variance of a polarization measurement. Since each photon is independent and $1/N$ is a constant factor, the variance can be expressed as a sum,
\begin{equation}
    \label{eq:Variance_Sum_Independent}
    \Var(P) = \frac{1}{N^2}\sum_{i=1}^N \Var(P_i).
\end{equation}

Using expectation values of the single-qubit polarization operator, we deduce that the variance of a single measurement of an ideal interferometer is
\begin{equation}
    \Var(P_i) = \langle P_i^2 \rangle - \langle P_i \rangle ^2 = 1 - \cos^2\phi = \sin^2 \phi.
\end{equation}
Here $\langle P_i^2 \rangle = 1$ because both polarization measurement outcomes square to 1, and $\langle P_i \rangle = -\cos{\phi}$ as derived in Eq.~\eqref{eq:polarization-definition}. Thus, $\Var(P) =  \sin^2 \phi/N$ and $\sigma_P = \sin \phi/\sqrt{N}$. Substituting back into Eq.~\eqref{eq:noise_slope}, the phase sensitivity is
\begin{equation}\label{eq:sigma_phi_1qubit_final}
    \sigma_\phi = \frac{1}{\sqrt{N}},
\end{equation}
which is equal to the standard quantum limit and is independent of $\phi$.

Figure~\ref{fig:1-qubit-sensitivity} shows how shot noise manifests in an interferometer implemented on a quantum computer. The central panel of the figure depicts the experimentally determined relationship between interferometer phase and polarization and a linear approximation of this curve around $\phi_0 = \pi/2$. This initial phase and the corresponding polarization are depicted as dashed lines in the figure. We illustrate the sensitivity of the interferometer to a phase shift $\delta\phi = 0.2$ by repeatedly measuring the polarization at the output of a quantum circuit where the interferometer phase was set to $\phi = \pi/2 + 0.2$. From each measurement of the polarization, we obtain an estimate $\tilde{\phi}$ using the previously determined relationship between the two variables. 

Repeatedly measuring polarization illustrates the uncertainty in each measurement due to shot noise and other technical noise sources. Each polarization measurement consists of $N=100$ shots. The histogram shown on the right margin is constructed from 75 independent measurements of probability to illustrate variability. The measured polarization has a standard deviation of 0.109(9), which is within error of theoretical value $\sigma_P = \sin{\phi}/\sqrt{N}$ at $\phi = \pi/2 + 0.2$. The distribution of polarization maps to a distribution of the inferred values of $\tilde{\phi}$ which has a mean of $\pi/2 + 0.20(1)$ and standard deviation of 0.117(10). The mean is consistent with the initially chosen $\delta\phi$, showing that $\tilde{\phi}$ is an unbiased estimator for the phase shifts. The width of the distribution of $\tilde{\phi}$ is the sensitivity of the interferometer to phase shifts. The estimated phase sensitivity of 0.117(10) is consistent with shot noise for $N=100$ shots combined with the factor of $1/m = 1.073$ due to the reduced amplitude of the polarization curve as compared to Eq.~\eqref{eq:noise_slope}. 

\begin{figure}[tp]
\centering
\includegraphics[width=0.45\textwidth]{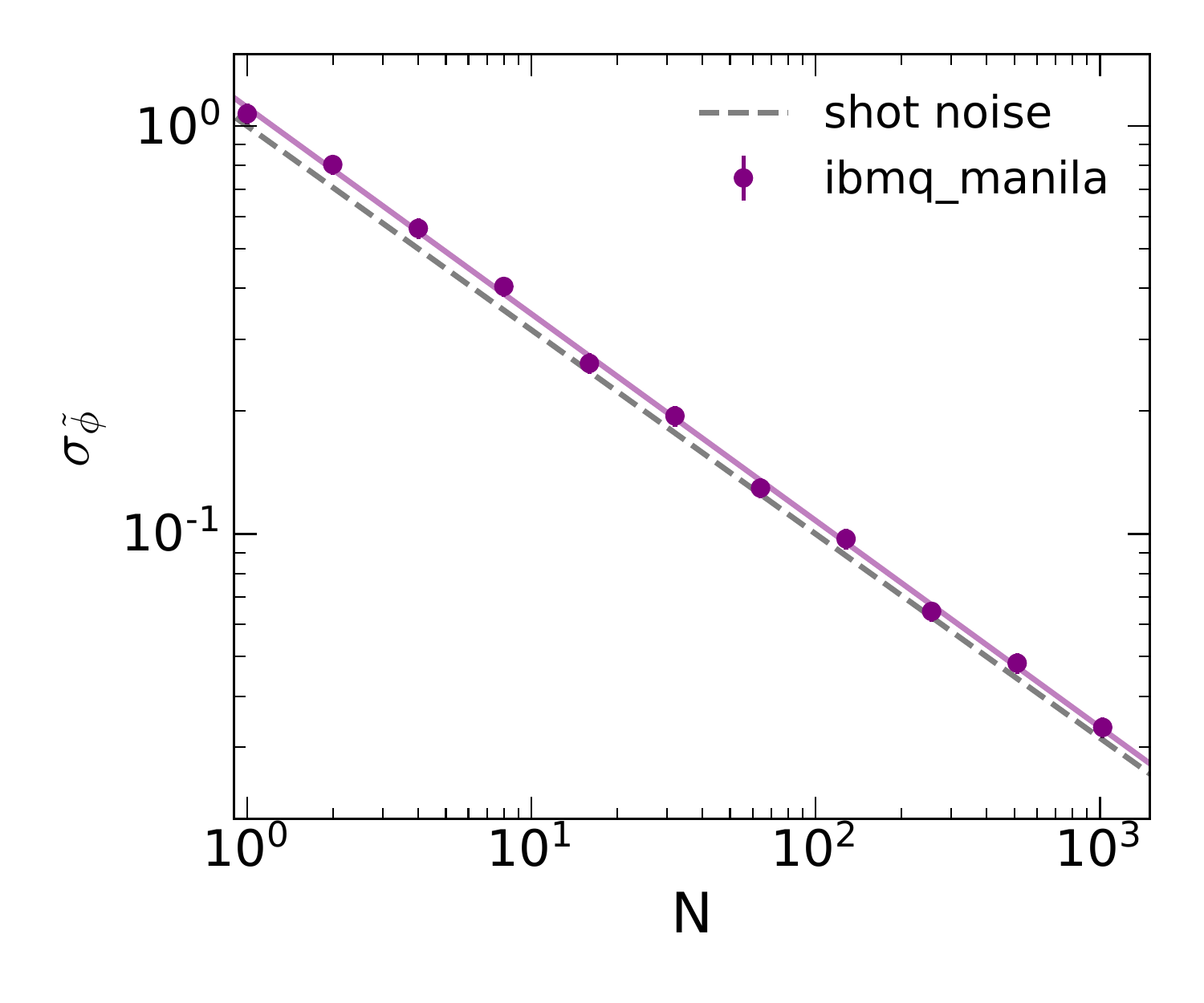}
\caption{Standard deviation of inferred $\tilde{\phi}$ from polarization data taken on ibmq\_manila as a function of number of qubits $N$. The least-squares best fit line for the log-log plot is shown in solid purple. All data points lie slightly above the standard quantum limit (dashed line) $1/\sqrt{N}$.}
\label{fig:1-qubit-shotnoise}
\end{figure}

The model we derived in Eq.~\eqref{eq:sigma_phi_1qubit_final} for the shot noise of independent photons predicts that the phase sensitivity scales with the number of shots $N$ as $\sigma_{\tilde{\phi}} \propto 1/\sqrt{N}$. To verify this scaling, we repeat the procedure performed in Fig.~\ref{fig:1-qubit-sensitivity} for values of $N$ between 1 and 1024. We plot the measured sensitivities $\sigma_{\tilde{\phi}}$ in Fig.~\ref{fig:1-qubit-shotnoise}. Shot noise corresponds to a slope of $-1/2$ on the log-log scale depicted in this figure. For all $N$, the measured sensitivities are slightly above shot noise, but follow the expected scaling, with a slope of -0.506(5) on the log-log plot. On average, the plotted sensitivities are 11(2)\% above shot noise, of which 7.3\% is due to the reduced amplitude of polarization curve from Fig.~\ref{fig:1-qubit_polarization}. The remaining deviation is not statistically significant, but may be due to weak correlations between successive shots, which violate the assumption of independence required for Eq.~\eqref{eq:Variance_Sum_Independent}. Overall, the results are consistent with the expected scaling for a predominantly shot noise limited device in the presence of some technical noise. The sensitivity of this interferometer is thus primarily limited by the number of qubits queried. This corresponds to noise scaling with the photons or optical power at LIGO.

\section{Modeling Quantum Enhanced Measurement}\label{Sec:2-qubit-result}
LIGO now utilizes entanglement to improve sensitivity without further increasing optical power. Specifically, LIGO now uses light in a squeezed quantum state where photons are entangled to each other~\cite{tse2019quantum}. The correlations between the photons in this state improve sensitivity to phase at the expense of amplitude uncertainty. Due to these correlations, the photons are not independent and can circumvent the standard quantum limit. This manipulation of quantum uncertainty is widely used in quantum metrology. Many examples of metrologically useful states can modeled using the language of quantum circuits~\cite{caves_2010_quantumcircuit}. 

The simplest form of entanglement involves two qubits. We can create a two-qubit entangled state on a quantum computer by using a combination of $H$, $X$, and CNOT gates. Specifically, the controlled not gate (CNOT) flips a target qubit (labeled with $\bigoplus$) based on the state of a control qubit (labeled with $\bullet$). This allows the two qubits to interact, producing an entangled state that can be used as the input to an interferometer circuit in order to achieve improved sensitivity~\cite{zhang2014quantum}. The full circuit is
\begin{equation}\label{eq:2-qubit-circuit}
    \Qcircuit @C=0.75em @R=0.2em @!R { \\
    	 	\nghost{ {q}_{0} :  } & \lstick{ {q}_{0} :  } & \gate{\mathrm{\left|0\right\rangle}} & \gate{\mathrm{H}} & \ctrl{1} & \gate{\mathrm{H}} & \gate{\mathrm{R_Z}\,(\mathrm{\phi})} & \gate{\mathrm{H}} & \meter & \qw & \qw & \qw\\ 
    	 	\nghost{ {q}_{1} :  } & \lstick{ {q}_{1} :  } & \gate{\mathrm{\left|0\right\rangle}} & \gate{\mathrm{X}} & \targ & \gate{\mathrm{H}} & \gate{\mathrm{R_Z}\,(\mathrm{\phi})} & \gate{\mathrm{H}} & \qw & \meter & \qw & \qw\\ 
    	 	\nghost{c:} & \lstick{c:} & \lstick{/_{_{2}}} \cw & \cw & \cw & \cw & \cw & \cw & \dstick{_{_{0}}} \cw \cwx[-2] & \dstick{_{_{1}}} \cw \cwx[-1] & \cw & \cw\\ 
    }
\end{equation}

To understand this circuit's dynamics, we first construct a mathematical representation of two-qubit states. An arbitrary two-qubit state can be expressed as a superposition of four two-qubit basis states $|\psi\rangle = c_0 |00\rangle+ c_1|01\rangle + c_2|10\rangle + c_3 |11\rangle$. The corresponding matrix notation is the $4 \times 1$ vector $[c_0, c_1, c_2, c_3]^T$. Here each basis state is the tensor product of single-qubit basis states for qubits $q_0$ and $q_1$. For example, $|01\rangle = |0\rangle_0 \otimes |1\rangle_1$. In general, states that can be represented as a tensor product $|\psi\rangle = |\varphi_0\rangle \otimes |\varphi_1\rangle$ are not entangled. If a two-qubit state cannot be factorized as a tensor product of single-qubit states, the two qubits are entangled~\cite{nielsen2010quantum}.

The first three gates in the two-qubit circuit (Eq.~\eqref{eq:2-qubit-circuit}) transform the initial state $|00\rangle$ into an entangled state $\frac{1}{\sqrt{2}} (|01\rangle + |10\rangle)$, known as a Fock state, to be used for quantum enhanced measurements. The first Hadamard gate and the $X$ gate act independently on the two qubits, preparing the product state $\frac{1}{\sqrt{2}}(|0\rangle_0 + |1\rangle_0) \otimes |1\rangle_1 = \frac{1}{\sqrt{2}} (|01\rangle + |11\rangle)$. Next, the CNOT gate produces entanglement by flipping the target bit $q_1$ when the control bit $q_0$ is in the state $|1\rangle$. This operation can be represented by the $4 \times 4$ matrix
\begin{equation}
    \text{CNOT} = \begin{bmatrix}
1 & 0 & 0 & 0 \\
0 & 1 & 0 & 0 \\
0 & 0 & 0 & 1 \\
0 & 0 & 1 & 0 \\
\end{bmatrix}.
\end{equation}
Applying this matrix to the state $\frac{1}{\sqrt{2}}[0, 1, 0, 1]^T$ produces the Fock state $\frac{1}{\sqrt{2}}[0, 1, 1, 0]^T = \frac{1}{\sqrt{2}} (|01\rangle + |10\rangle)$.

\begin{figure*}[tp]
    \centering
    \includegraphics[width=0.75\textwidth]{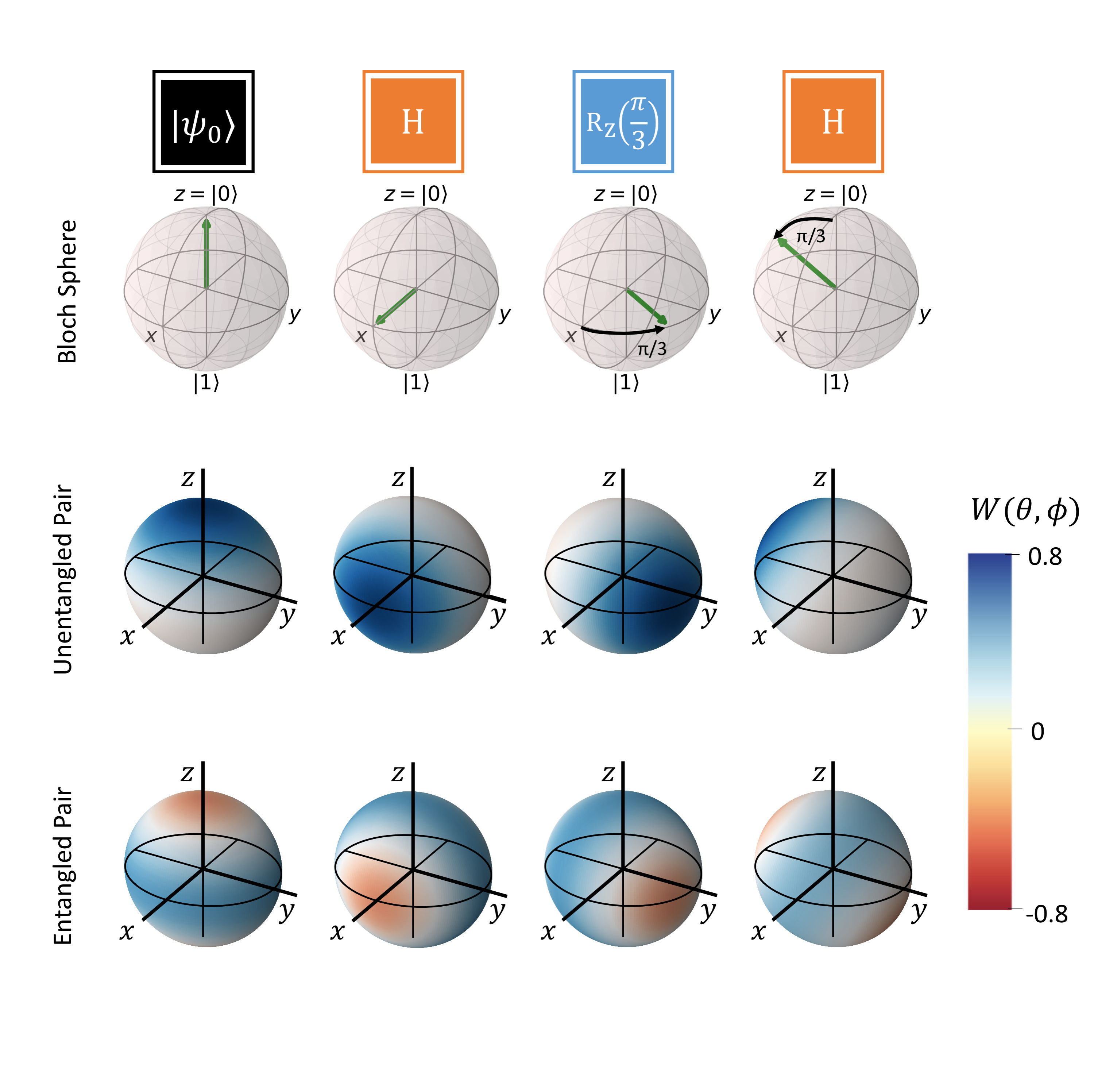}
    \caption{Visualization of the quantum state over time in the two-qubit circuit using the spin Wigner distribution $W(\theta,\phi)$. Successive columns depict evolution of the quantum state of the system after each gate in the interferometry sequence is applied. The first row reproduces the single-qubit Bloch spheres from Fig.~\ref{fig:Bloch-sphere-1-qubit} for comparison. The second row depicts the Wigner distribution for an unentangled initial state $|\psi_0\rangle = |00\rangle$, which is localized about the single-qubit Bloch vector. The third row depicts the Wigner distribution for an entangled pair of qubits initialized in the Fock state $|\psi_0\rangle = \frac{1}{\sqrt{2}}(|01\rangle + |10\rangle)$, appearing as a horizontal ring. The shape of the Wigner quasi-probability distribution illustrates measurement uncertainty for each quantum state and is rotated under the application of gates in the interferometer sequence in same way as the Bloch vector.}
    \label{fig:Bloch-sphere-2-qubit}
\end{figure*}

The value of using an entangled state like the Fock state for quantum metrology can be understood visually through the Wigner representation on a spin sphere~\cite{dowling1994wigner}. The Wigner function is a quasi-probability distribution that maps the uncertainty of the quantum state onto the spin sphere. Similar to the Bloch sphere representation, single-qubit gates applied to all qubits rotate the Wigner function without changing the shape of the distribution. This allows for a clean representation of the behavior of each state as it evolves through the interferometer circuit. Figure~\ref{fig:Bloch-sphere-2-qubit} shows the evolution of the two-qubit state after each gate using the Wigner distribution. The single-qubit Bloch spheres from Fig.~\ref{fig:Bloch-sphere-1-qubit} are reproduced in the first row for comparison.

The second and third row of Fig.~\ref{fig:Bloch-sphere-2-qubit} compare the evolution of the Wigner distribution when the two qubits start in an untangled state $|00\rangle$ and when they start in the Fock state $\frac{1}{\sqrt{2}}(|01\rangle + |10\rangle)$, respectively. The Wigner function of the unentangled pair is a localized distribution centered about the Bloch vector, with equal uncertainty in both x and y directions. The width of this local distribution (dark blue region) on the unit sphere is $1/\sqrt{2}$ radians~\cite{dowling1994wigner}, corresponding to the shot noise of two unentangled qubits. For an entangled state like the Fock state, the Wigner distribution is not necessarily symmetric and can have less uncertainty in one direction. The Fock state Wigner function appears as a horizontal ring centered about z=0 in the bottom left sphere in Fig.~\ref{fig:Bloch-sphere-2-qubit}. The thickness of the ring, denoting the uncertainty in the z-direction, is smaller than the width of the unentangled Wigner distribution, at the expense of increased uncertainty in the azimuthal direction. Entanglement also leads to negative Wigner function values at the poles. When properly rotated during the interferometer sequence, this entangled state can exhibit increased sensitivity to the interferometer phase $\phi$ due to the reduced width and increased spatial structure of Wigner distribution.

To precisely model the behavior of the Fock state in the interferometer requires matrix representations of the three gates used in the interferometer sequence. Since these gates act independently on the two qubits, we can represent each step of the interferometer as the tensor product of two single-qubit gates. The interferometer's beam splitter is represented by a pair of Hadamard gates,
\begin{equation}
    H \otimes H = \frac{1}{2}\begin{bmatrix}
1 & 1 & 1 & 1 \\
1 & -1 & 1 & -1 \\
1 & 1 & -1 & -1 \\
1 & -1 & -1 & 1 \\
\end{bmatrix}.
\end{equation}
The application of this Hadamard gate transforms the Fock state $\frac{1}{\sqrt{2}}(|01\rangle + |10\rangle)$ into a cat state $\frac{1}{\sqrt{2}}(|00\rangle - |11\rangle)$. Graphically this corresponds to rotating the Wigner distribution by $\pi/2$ such that the ring is oriented vertically and is sensitive to azimuthal rotations due the interferometer phase.

The phase shift portion on the interferometer is given by the product of individual qubit rotations,
\begin{equation}
R_Z(\phi) \otimes R_Z(\phi) = \begin{bmatrix}
e^{-i\phi} & 0 & 0 & 0 \\
0 & 1 & 0 & 0 \\
0 & 0 & 1 & 0 \\
0 & 0 & 0 & e^{i\phi} \\
\end{bmatrix}.
\end{equation}
The application of this gate to the cat state results in the state $|\psi\rangle=e^{-i\phi}|00\rangle - e^{i\phi} |11\rangle$, introducing a phase shift of $2\phi$ between the basis states. The state oscillates as a function of $\phi$ between the two cat states $\frac{1}{\sqrt{2}}(|00\rangle - |11\rangle)$ and $\frac{1}{\sqrt{2}}(|00\rangle + |11\rangle)$ with twice the frequency at which states in the single-qubit interferometer oscillate around the Bloch sphere. 

The final pair of Hadamard gates represents the second pass through a beam splitter and maps the phase shift from the $R_Z$ gates onto the computation basis. The Hadamard gates map one cat state $\frac{1}{\sqrt{2}}(|00\rangle - |11\rangle)$ back to the initial Fock state, while leaving the other cat state $\frac{1}{\sqrt{2}}(|00\rangle + |11\rangle)$ unchanged. Thus the final state is an oscillation between the Fock state and the cat state, 
\begin{equation}\label{eq:final_state_2qubit}
|\psi\rangle = \frac{1}{2\sqrt{2}}
\begin{bmatrix}
e^{-i\phi} - e^{i\phi} \\
e^{-i\phi} + e^{i\phi} \\
e^{-i\phi} + e^{i\phi} \\
e^{-i\phi} - e^{i\phi}
\end{bmatrix}
= \frac{\cos{\phi}}{\sqrt{2}} \begin{bmatrix}
0 \\
1 \\
1 \\
0
\end{bmatrix} - \frac{i\sin{\phi}}{\sqrt{2}}\begin{bmatrix}
1 \\
0 \\
0 \\
1
\end{bmatrix}.
\end{equation}
The Fock state is detected when the measurements of the two qubits are different, which has a total probability of $p_{01} + p_{10} = \cos^2{\phi}$. The cat state corresponds to both qubits being in the same state, which has a total probability of $p_{00} + p_{11} = \sin^2{\phi}$.

To encapsulate the oscillation between the cat state and Fock state in a single number, we define a parity operator such that parity is 1 when both qubits are in the same state (e.g., cat state), and parity is -1 when the qubits are in different states (e.g., Fock state). In matrix form, this parity operator is 
\begin{equation}\label{eq:parity_matrix}
    \hat{\Pi} = \begin{bmatrix}
    1 & 0 & 0 & 0 \\
    0 & -1 & 0 & 0 \\
    0 & 0 & -1 & 0 \\
    0 & 0 & 0 & 1
    \end{bmatrix}.
\end{equation}
The expectation value of the operator is
\begin{equation}\label{eq:two_qubit_parity}
\begin{aligned}
     \langle \Pi \rangle &= \langle \psi |\hat{\Pi}| \psi \rangle \\ 
     &= (p_{00} + p_{11}) - (p_{01} + p_{10}) \\
     &= -\cos{(2\phi)},
     \end{aligned}
\end{equation}
in close analogy to the polarization measurement for a single-qubit interferometer. 

The key difference between the expectation value of the two-qubit parity measurement (Eq.~\eqref{eq:two_qubit_parity}) and the expectation value of the average polarization of independent qubits (Eq.~\eqref{eq:polarization-definition}) is the frequency of oscillation when $\phi$ is varied continuously. As illustrated by Wigner distributions in Fig.~\ref{fig:Bloch-sphere-2-qubit}, an untangled state returns to its original orientation after a rotation of $\phi = 2\pi$ whereas the ring-shaped Wigner distribution for an entangled pair of qubits is indistinguishable from its initial state after a rotation by $\phi=\pi$. The increased frequency of oscillation leads to an increased slope $m=2\sin(2\phi)$ of the parity vs $\phi$ curve as compared to the slope $m=\sin(\phi)$ of the polarization curve for the single-qubit interferometer. This improves phase sensitivity. 

In analogy to Eq.~\eqref{eq:noise_slope}, the phase sensitivity achievable by parity measurement is
\begin{equation}\label{eq:2-qubit-sigma-phi}
    \sigma_{\tilde{\phi}} = \frac{\sigma_\Pi}{2\sin{(2\phi)}},
\end{equation}
where $\sigma_{\Pi}$ is the standard deviation of the average parity. The standard deviation of the average parity in measurement consisting of $n$ shots of the two-qubit interferometer can be computed using the matrix form of the single-shot parity operator $\hat{\Pi}_i$. Assuming all shots are independent,
\begin{equation}
    \sigma_{\Pi}^2 =  \frac{\Var({\Pi_i})}{n} = \frac{\langle \hat{\Pi}_i^2 \rangle - \langle \hat{\Pi}_i \rangle^2}{n} = \frac{\sin^2{(2\phi)}}{n}.
\end{equation}
For a fair comparison to the single-qubit interferometer, we note that $n$ shots involve measurements of $N=2n$ qubits so the properly normalized phase sensitivity is
\begin{equation}\label{eq:2-qubit-phase-sensitivity}
    \sigma_{\tilde{\phi}} = \frac{1}{\sqrt{2N}}.
\end{equation}
This is an overall improvement by a factor of $\sqrt{2}$ as compared to the shot noise in an unentangled system.

\section{Quantifying Benefits of Entanglement}\label{Sec:2-qubit-noise}
\begin{figure}[tp]
\centering
\includegraphics[width=0.48\textwidth]{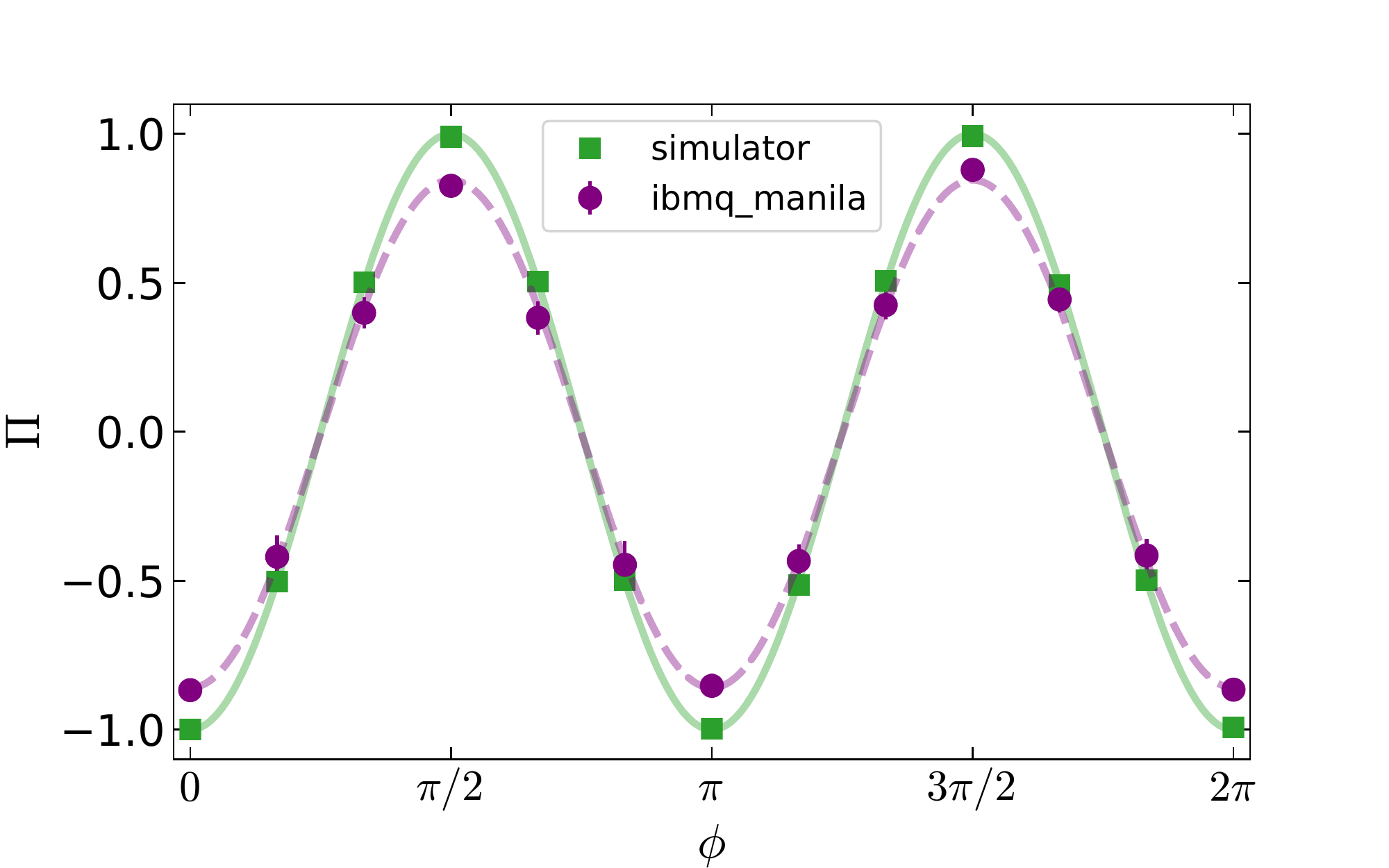}
\caption{Measured parity, $\Pi = p_{00} + p_{11} - p_{01} - p_{10}$, as a function of $\phi$. Purple circles show data from the two-qubit interferometer taken on the quantum computer ibmq\_manila at 13 angles between 0 and $2\pi$. Green squares show data from the simulator ibmq\_qasm\_simulator at the same angles.  Error bars depict the standard deviation of 5 trials. The curves are fit to the data with amplitude and offset as free parameters.}
\label{fig:2-qubit_manila_parity}
\end{figure}

To demonstrate the improvement in sensitivity introduced by entanglement, we implemented the two-qubit interferometer circuit on the same quantum computer, ibmq\_manila, that was used earlier to implement the single-qubit interferometer. We additionally implemented the circuit on the state vector simulator, ibmq\_qasm\_simulator, to provide a comparison without the effects of technical noise. Figure~\ref{fig:2-qubit_manila_parity} shows the results of parity measurements for each of the two implementations. Each measurement of parity $\Pi = p_{00} + p_{11} - p_{01} - p_{10}$ is computed from $n = 1024$ shots of the 2-qubit interferometer circuit. The results show the expected sinusoidal dependence on $\phi$ with a frequency twice that of the single-qubit interferometer.

We fit the Manila and simulator data to the expected functional form $\Pi = -A\cos{(2\phi)}+B$ with amplitude and vertical offset as free parameters. The simulator fit has an offset of -0.001(2) and amplitude of 0.997(3), both of which agree well with the expected values of 0 and 1. The Manila fit's offset of -0.007(6) is also consistent with 0, while the amplitude of 0.852(7) corresponds to a 15\% reduction from the ideal value. This deviation in amplitude is larger than the corresponding value for the single-qubit interferometer (6.8\%) because readout and gate errors in either qubit can flip overall parity, doubling the contribution from these errors. Additionally, the CNOT gate has an error of up to 1\%~\cite{IBM_compute}, (varying between daily calibrations), which further contributes to the reduction in amplitude. 

\begin{figure}[tp]
\centering
\includegraphics[width=0.42\textwidth]{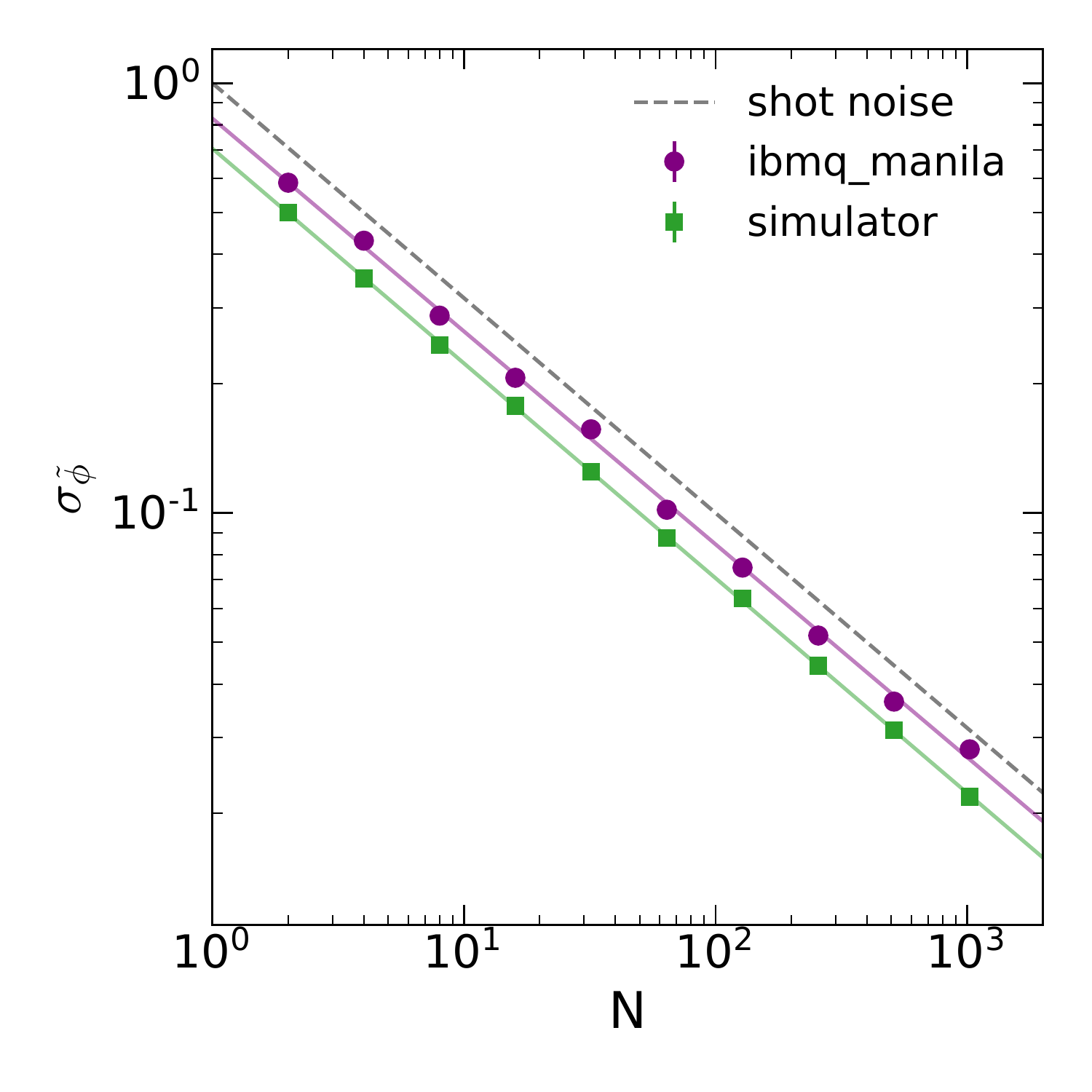}
\caption{Standard deviation of inferred $\tilde{\phi}$ from polarization data taken on the quantum computer Manila and the state vector simulator as a function of number of qubits $N$ along with their best-fit lines. All data points lie below the standard quantum limit (dashed line) $1/\sqrt{N}$.}
\label{fig:2-qubit-shotnoise}
\end{figure}

Curves of parity $\Pi$ vs phase $\phi$ can be used to infer phase $\tilde{\phi}$ from measurements of the two-qubit, in analogy to the procedure illustrated for single-qubit polarization measurements in Fig.~\ref{fig:1-qubit-sensitivity} and Fig.~\ref{fig:1-qubit-shotnoise}. To analyze the phase sensitivity quantitatively, like we did for the one-qubit interferometer, we collected parity data on both the quantum computer, ibmq\_manila, and the simulator. Each measurement of parity came from a trial consisting of 1 to 512 sequential shots (corresponding to $N=$ 2 to 1024 qubits), conducted with an interferometer phase $\phi = \pi/4$. From each parity measurement, we inferred a value for the interferometer phase $\tilde{\phi}$ using the slope of the cosine fits from Fig.~\ref{fig:2-qubit_manila_parity}. We computed the standard deviation of these phase measurements $\sigma_{\tilde{\phi}}$ using data from 600 trials. The resulting relationship between phase uncertainty and number of qubits contributing to each measurement is plotted in Fig.~\ref{fig:2-qubit-shotnoise}.

The measured phase sensitivities for both ibmq\_manila and simulator data are consistently below the standard quantum limit $1/\sqrt{N}$ for unentangled qubits. Fitting the simulator data verifies that the phase sensitivity is $1/\sqrt{2N}$, as expected for $N/2$ pairs of entangled qubits. The fit for measurements made on the quantum computer is $(0.83 \pm 0.02) N^{-0.495(6)}$, which lies between the standard quantum limit and the limit for entangled pairs of qubits. Overall, the two-qubit interferometer shows an improvement of 17(2)\% as compared to the standard quantum limit and 25(3)\% when compared to the single-qubit interferometer implemented on the same hardware. This illustrates the benefits that entanglement can provide for quantum measurement, even in the smallest possible system sizes. 

\section{Conclusions}\label{Sec:conclusions}
We have demonstrated how entanglement between two qubits enables measurement with sensitivity below the standard quantum limit. Students working on these experiments gain better comprehension of quantum mechanics, quantum metrology, and entanglement. Running these experiments on a real quantum computer, students practice quantum programming, analyze experimental data, and assess the effects of both statistical uncertainty and technical noise sources. This prepares students with the tools needed to engage in modern research in quantum science and technology.

The project presented here requires minimal prerequisites. The current work began as an independent student project through a STEM outreach program that pairs high school student researchers with graduate student mentors~\cite{galvin2021curiosity}. This illustrates how the material can be accessible to motivated students for whom this is their first introduction to quantum mechanics. We believe this project would fit well in the undergraduate curriculum, either as a demonstration for an introductory quantum information course or as a project in a modern physics laboratory class. These experiments can be accomplished with minimal resources as the IBM Quantum Experience allows for free remote access. Additionally, students working independently for a class project could extend the techniques presented here to circuits with more than two qubits~\cite{encinar2021digital} in order to demonstrate different forms of entanglement and implement a variety of quantum squeezing techniques.




\begin{acknowledgments}
We thank Stanford FAST for facilitating our collaboration and providing support for this research. TN acknowledges support from NSF Q-SEnSE and ESC acknowledges the support of the NSF Graduate Research Fellowship. We additionally thank Victoria Borish, Monika Schleier-Smith and Patrick Allamandola for useful discussions. 
\end{acknowledgments}

\section*{Author contributions}
All authors contributed equally to this project. CT conducted experiments with guidance from TN and ESC. All authors participated in the analysis of experimental data and the preparation of this manuscript.


\begin{thebibliography}{99}
\bibitem{brody2021spin} J.~Brody and G.~Guzman, ``Calculating spin correlations with a quantum computer.'' Am. J. Phys. \textbf{89}, 35 (2021). 

\bibitem{johnstuna2021understanding} S.~Johnstuna and J.-F. Van Hueleb, ``Understanding and compensating for noise on IBM quantum computers'' Am.~J.~Phys.~\textbf{89}, 935 (2021) 

\bibitem{kain2021searching} B.~Kain, ``Searching a quantum database with Grover's search algorithm,'' Am.~J.~Phys.~\textbf{89}, 618 (2021)

\bibitem{encinar2021digital} P.~C.~Encinar, A.~Agustí, and C.~Sabín, ``Digital quantum simulation of beam splitters and squeezing with IBM quantum computers,'' Physical Review A 104.5 (2021): 052609.

\bibitem{clocks} L.~I.~R. Gil, R. Mukherjee, E.~M. Bridge, M.~P.~A. Jones, and T. Pohl, ``Spin Squeezing in a Rydberg Lattice Clock,'' Phys.~Rev.~Lett.~\textbf{112}, 103601 (2014)

\bibitem{bio_imaging} Y.~Israel, S.~Rosen, and Y.~Silberberg, ``Supersensitive Polarization Microscopy Using NOON States of Light,''
Phys.~Rev.~Lett.~\textbf{112}, 103604 (2014)

\bibitem{remote_sensing} S.~R. Zhao \textit{et al.}, ``Field Demonstration of Distributed Quantum Sensing without Post-Selection,''
Phys.~Rev.~X~\textbf{11}, 031009 (2021)

\bibitem{abbot2016observation} B.~P.~Abbott \textit{et al.} (LIGO Scientific Collaboration and Virgo Collaboration), ``Observation of Gravitational Waves from a Binary Black Hole Merger,''
Phys.~Rev.~Lett.~\textbf{116}, 061102 (2016).

\bibitem{tse2019quantum} M.~Tse \textit{et al.}, ``Quantum-Enhanced Advanced LIGO Detectors in the Era of Gravitational-Wave Astronomy,'' Phys.~Rev.~Lett.~\textbf{123}, 231107 (2019).

\bibitem{georgescu2020o3} I.~Georgescu, ``O3 highlights,'' Nat.~Rev.~Phys.~\textbf{2}, 222–223 (2020).

\bibitem{cooper2021interactive} S.~J.~Cooper \textit{et al.}, ``An interactive gravitational-wave detector model for museums and fairs,'' Am. J. Phys. \textbf{89}, 702 (2021).

\bibitem{dowling1994wigner} J.~P.~Dowling, G.~S.~Agarwal, and W.~P.~Schleich, ``Wigner distribution of a general angular-momentum state: Applications to a collection of two-level atoms,'' Phys. Rev. A \textbf{49}, 4101 (1994)

\bibitem{abbott2020noise} B.~P.~Abbott et al, ``A guide to LIGO–Virgo detector noise and extraction of transient gravitational-wave signals,'' Class. Quantum Grav. \textbf{37}, 055002 (2020)

\bibitem{caves_2010_quantumcircuit} C.~M.~Caves, A.~Shaji, ``Quantum-circuit guide to optical and atomic interferometry,'' Opt. Commun. \textbf{283} (5), 695-712 (2010)

\bibitem{operation} ``LIGO's Interferometer,'' Caltech, \url{https://www.ligo.caltech.edu/page/ligos-ifo/} (2021)

\bibitem{nielsen2010quantum} M.~A.~Nielsen and I.~L.~Chuang, ``Quantum Computation and Quantum Information,'' Cambridge University Press, Cambridge, 2010, pp. 290-296.

\bibitem{asfaw2020} A.~Asfaw \textit{et al.}, ``Learn Quantum Computation Using Qiskit,'' \url{ http://community.qiskit.org/textbook} (2020)

\bibitem{github} ``LIGO-code'' \url{ https://github.com/cindytrann/quLIGO-code.git} (2021)

\bibitem{IBM_compute} IBM Quantum, ``Compute resources'' \url{https://quantum-computing.ibm.com/services/resources}

\bibitem{quantum_volume} A.~W.~Cross \textit{et al.}, ``Validating quantum computers using randomized model circuits,'' Phys. Rev. A \textbf{100}, 032328 (2019)

\bibitem{zhang2014quantum} Z.~Zhang and L.~M.~Duan, ``Quantum metrology with Dicke squeezed states.'' New J. Phys. \textbf{16} 103037 (2014)

\bibitem{galvin2021curiosity} C.~J.~Galvin \textit{et al.}, ``Curiosity-Based Biophysics Projects in a High School Setting with Graduate Student Mentorship,'' The Biophysicist \textbf{2} (1): 6–11 (2021)

\end{thebibliography}
\end{document}